\documentclass[twoside,twocolumn,9pt]{article}
\usepackage{extsizes}
\usepackage[super,sort&compress,comma]{natbib} 
\usepackage[version=3]{mhchem}
\usepackage[left=1.5cm, right=1.5cm, top=1.785cm, bottom=2.0cm]{geometry}
\usepackage{balance}
\usepackage{mathptmx}
\usepackage{sectsty}
\usepackage{graphicx} 
\usepackage{lastpage}
\usepackage[format=plain,justification=justified,singlelinecheck=false,font={stretch=1.125,small,sf},labelfont=bf,labelsep=space]{caption}
\usepackage{float}
\usepackage{fancyhdr}
\usepackage{fnpos}
\usepackage[english]{babel}
\addto{\captionsenglish}{%
  
}
\usepackage{array}
\usepackage{droidsans}
\usepackage{charter}
\usepackage[T1]{fontenc}
\usepackage[usenames,dvipsnames]{xcolor}
\usepackage{setspace}
\usepackage[compact]{titlesec}
\usepackage{hyperref}

\usepackage{amsmath,amssymb}
\usepackage{upgreek}

\usepackage{epstopdf}

\definecolor{cream}{RGB}{222,217,201}

\begin{document}

\pagestyle{fancy}
\thispagestyle{plain}
\fancypagestyle{plain}{
\renewcommand{\headrulewidth}{0pt}
}

\makeFNbottom
\makeatletter
\renewcommand\LARGE{\@setfontsize\LARGE{15pt}{17}}
\renewcommand\Large{\@setfontsize\Large{12pt}{14}}
\renewcommand\large{\@setfontsize\large{10pt}{12}}
\renewcommand\footnotesize{\@setfontsize\footnotesize{7pt}{10}}
\makeatother

\renewcommand{\thefootnote}{\fnsymbol{footnote}}
\renewcommand\footnoterule{\vspace*{1pt}%
\color{cream}\hrule width 3.5in height 0.4pt \color{black}\vspace*{5pt}} 
\setcounter{secnumdepth}{5}

\makeatletter 
\renewcommand\@biblabel[1]{#1}            
\renewcommand\@makefntext[1]%
{\noindent\makebox[0pt][r]{\@thefnmark\,}#1}
\makeatother 
\renewcommand{\figurename}{\small{Fig.}~}
\sectionfont{\sffamily\Large}
\subsectionfont{\normalsize}
\subsubsectionfont{\bf}
\setstretch{1.125} 
\setlength{\skip\footins}{0.8cm}
\setlength{\footnotesep}{0.25cm}
\setlength{\jot}{10pt}
\titlespacing*{\section}{0pt}{4pt}{4pt}
\titlespacing*{\subsection}{0pt}{15pt}{1pt}

\fancyfoot{}
\fancyfoot[LO,RE]{\vspace{-7.1pt}\includegraphics[height=9pt]{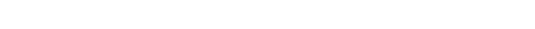}}
\fancyfoot[CO]{\vspace{-7.1pt}\hspace{13.2cm}\includegraphics{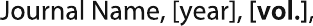}}
\fancyfoot[CE]{\vspace{-7.2pt}\hspace{-14.2cm}\includegraphics{head_foot/RF}}
\fancyfoot[RO]{\footnotesize{\sffamily{1--\pageref{LastPage} ~\textbar  \hspace{2pt}\thepage}}}
\fancyfoot[LE]{\footnotesize{\sffamily{\thepage~\textbar\hspace{3.45cm} 1--\pageref{LastPage}}}}
\fancyhead{}
\renewcommand{\headrulewidth}{0pt} 
\renewcommand{\footrulewidth}{0pt}
\setlength{\arrayrulewidth}{1pt}
\setlength{\columnsep}{6.5mm}
\setlength\bibsep{1pt}

\makeatletter 
\newlength{\figrulesep} 
\setlength{\figrulesep}{0.5\textfloatsep} 

\newcommand{\topfigrule}{\vspace*{-1pt}%
\noindent{\color{cream}\rule[-\figrulesep]{\columnwidth}{1.5pt}} }

\newcommand{\botfigrule}{\vspace*{-2pt}%
\noindent{\color{cream}\rule[\figrulesep]{\columnwidth}{1.5pt}} }

\newcommand{\dblfigrule}{\vspace*{-1pt}%
\noindent{\color{cream}\rule[-\figrulesep]{\textwidth}{1.5pt}} }

\makeatother

\twocolumn[
  \begin{@twocolumnfalse}
{\includegraphics[height=30pt]{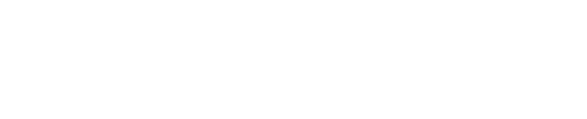}\hfill\raisebox{0pt}[0pt][0pt]{\includegraphics[height=55pt]{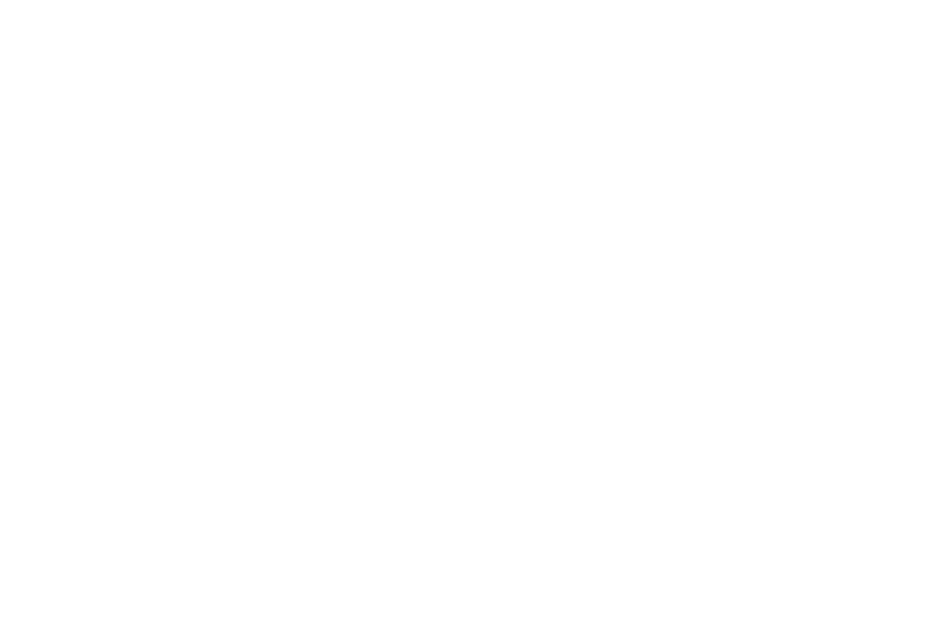}}\\[1ex]
\includegraphics[width=18.5cm]{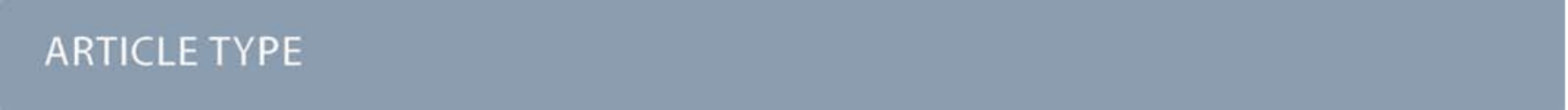}}\par
\vspace{1em}
\sffamily
\begin{tabular}{m{4.5cm} p{13.5cm} }

\includegraphics{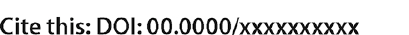} & \noindent\LARGE{\textbf{Anisotropic dynamics of a self-assembled colloidal chain in an active bath$^\dag$}} \\
\vspace{0.3cm} & \vspace{0.3cm} \\

  & \noindent\large{Mehdi Shafiei Aporvari,$^{\ast}$\textit{$^{ab}$} Mustafa Utkur,\textit{$^{bc}$} Emine Ulku Saritas,\textit{$^{bc}$} Giovanni Volpe,\textit{$^{d}$} and Joakim Stenhammar $^{\ast}$\textit{$^{e}$}} \\

\includegraphics{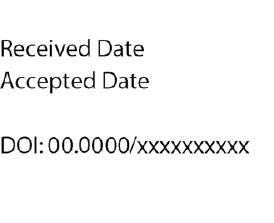} & \noindent\normalsize{Anisotropic macromolecules exposed to non-equilibrium (active) noise are very common in biological systems, and an accurate understanding of their anisotropic dynamics is therefore crucial. Here, we experimentally investigate the dynamics of isolated chains assembled from magnetic microparticles at a liquid-air interface and moving in an active bath consisting of motile \emph{E. coli} bacteria. We investigate both the internal chain dynamics and the anisotropic center-of-mass dynamics through particle tracking. We find that both the internal and center-of-mass dynamics are greatly enhanced compared to the passive case, \emph{i.e.}, a system without bacteria, and that the center-of-mass diffusion coefficient $D$ features a non-monotonic dependence as a function of the chain length. Furthermore, our results show that the relationship between the components of $D$ parallel and perpendicular with respect to the direction of the applied magnetic field is preserved in the active bath compared to the passive case, with a higher diffusion in the parallel direction, in contrast to previous findings in the literature. We argue that this qualitative difference is due to subtle differences in the experimental geometry and conditions and the relative roles played by long-range hydrodynamic interactions and short-range collisions.} \\

\end{tabular}

 \end{@twocolumnfalse} \vspace{0.6cm}

  ]

\renewcommand*\rmdefault{bch}\normalfont\upshape
\rmfamily
\section*{}
\vspace{-1cm}


\footnotetext{\textit{$^{a}$~UNAM -- National Nanotechnology Research Center, Bilkent University, Ankara 06800, Turkey. E-mail: mehdi.shafiei@bilkent.edu.tr}}
\footnotetext{\textit{$^{b}$~National Magnetic Resonance Research Center (UMRAM), Bilkent University, Ankara 06800, Turkey. }}

\footnotetext{\textit{$^{c}$~Department of Electrical and Electronics Engineering, Bilkent University, Ankara, Turkey. }}
\footnotetext{\textit{$^{d}$~Department of Physics, University of Gothenburg, SE-41296 Gothenburg, Sweden. }}
\footnotetext{\textit{$^{e}$~Division of Physical Chemistry, Lund University, Box 124, S-221 00 Lund, Sweden. E-mail: joakim.stenhammar@fkem1.lu.se }}

\footnotetext{\dag~Electronic Supplementary Information (ESI) available: [Series of videos demonstrating the motion of self-assembled  chains in active and passive bath]. See DOI: 00.0000/00000000.}


\section{introduction}

Chain-like biological objects with internal degrees of freedom are ubiquitous in nature: important examples include DNA molecules \citep{smith1995self, perkins1994relaxation}, actin filaments \citep{li2004diffusion}, intracellular microtubules \citep{le2002tracer,glaser2010tube, mitchison1984dynamic, brangwynne2008nonequilibrium}, and flagellar filaments \citep{magariyama1995simultaneous}. All of these systems naturally occur in environments far from thermodynamic equilibrium, where their statistical description requires going beyond the classical descriptions of equilibrium polymer theory. These facts have motivated a significant recent interest in both ``active polymers'', whose monomers are internally driven out of equilibrium through self-propulsion, and passive polymers in an ``active bath'', \emph{i.e.}, an environment containing self-propelled particles \cite{harder2014activity, eisenstecken2017internal, chaki2019enhanced, eisenstecken2016conformational, mallory2014curvature, winkler2017active}. These studies have revealed significant differences in polymer dynamics compared to the corresponding passive systems: ratchet currents and the emergence of nonuniform pressures along curved filaments~\cite{nikola2016active}, anomalous polymer swelling~\cite{kaiser2014unusual,kaiser2015does,vandebroek2015dynamics}, non-monotonic dependence of diffusivity on chain length~\cite{shin2017elasticity}, and facilitation of polymer looping~\cite{shin2015facilitation}. Most of these studies considered ``dry'' computational models based on so-called active Brownian particles \cite{volpe2014simulation}, where active and passive particles interact through short-ranged collisions, thus ignoring fluid-mediated hydrodynamic interactions that often play an important role in active matter systems \cite{bechinger2016active}. 

\begin{figure}
\includegraphics[width=1\columnwidth]{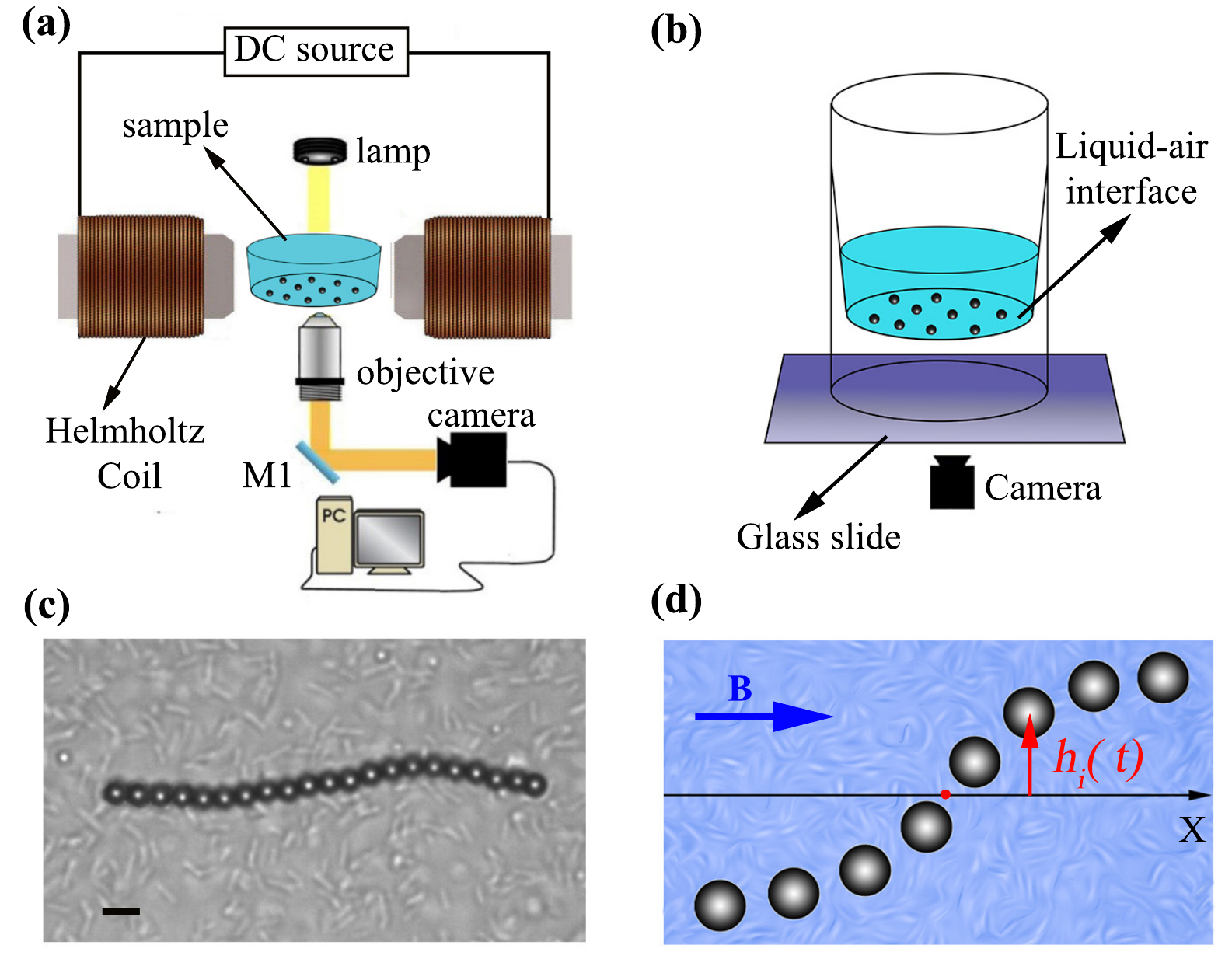}
\caption{Schematics of (a) the experimental setup and (b) the sample cell used to study colloidal particles at a liquid-air interface. The size of the sample cell in (a) and (b) has been exaggerated for clarity. (c) Bright-field microscopy image of a magnetic chain trapped at the liquid-air interface in a bacterial bath (scale bar: $5~{\rm \upmu m}$) and (d) sketch of the chain with the coordinate system used for the analysis of the internal chain dynamics. The red dot indicates the position of the center of mass.}
\label{fig1}
\end{figure}

Whether hydrodynamic or collisional in origin, the theoretical framework for understanding the effect of an active surrounding on suspended passive particles has often started from the viewpoint of equilibrium concepts such as temperature and free energy \cite{wu2000particle, leptos2009dynamics, mino2011enhanced, argun2016non}. For example, the strongly enhanced diffusion coefficient $D$ of spherical tracer particles in bacterial and algal baths can at long times be quantified through an elevated ``effective temperature'' defined through the Einstein relation $D = k_{\rm B} T_{\rm eff} / \gamma$, where $k_{\rm B}$ is Boltzmann's constant and $\gamma$ is the friction coefficient of the particle \cite{wu2000particle, leptos2009dynamics, mino2011enhanced}. However, in non-homogeneous environments, such as in the presence of confining potentials, this effective temperature description breaks down qualitatively, and the intrinsically non-equilibrium properties of the system become evident \cite{argun2016non}. Furthermore, the simplified effective temperature description often leads to different values of $T_{\rm eff}$ for the same system, depending on what observable one uses for its definition~\cite{loi2011non}. Thus, such simplified, effective thermodynamic descriptions should be used with care, although they are often useful as tools for qualitatively analysing experimental data.

In this article, we study a simple experimental realisation of a passive anisotropic body in an active bath: a chain of $N$ magnetic spheres moving in 2 dimensions trapped at a liquid-air interface, inside a bath of swimming \emph{E. coli} bacteria that interact with the chain through hydrodynamic as well as steric forces (Figs.~\ref{fig1}a-b).
For large magnitudes of the applied magnetic field $\mathbf{B}$, the strong induced dipolar forces between the particles lead to a stiff, rod-like chain aligned with the field direction, while for lower fields, the larger chain flexibility allows for significant shape fluctuations. We show that the internal as well as the center-of-mass dynamics of the chain are strongly enhanced, by an order of magnitude or more, in the active bath compared to the corresponding dynamics in the passive bath, i.e., a suspension without bacteria. We find that the ratio $D_{\parallel} / D_{\perp}$ between the components of the chain diffusion coefficient parallel and perpendicular to $\mathbf{B}$, is larger than in the corresponding passive case, where we find that $D_{\parallel} / D_{\perp} \approx 2$ for long chains, in accordance with the theoretical prediction for a long rigid chain in a 3-dimesional bulk system \citep{dhont1996introduction}.
This strongly contrasts with recent experimental findings using anisotropic tracer particles in \emph{E. coli} suspensions that found this ratio to change drastically, and even fall below unity, as the bacterial concentration was increased \cite{peng2016diffusion}. We argue that this qualitative difference is due to differences in the  intricate balance between short-ranged steric and long-ranged hydrodynamic forces in these different systems, indicating that the dynamics of tracer particles in active suspensions depend sensitively on the details of the experimental setup and the balance between hydrodynamic, steric and thermal forces.

\section{Experimental details}

\begin{figure*}
\includegraphics[width=2.0\columnwidth]{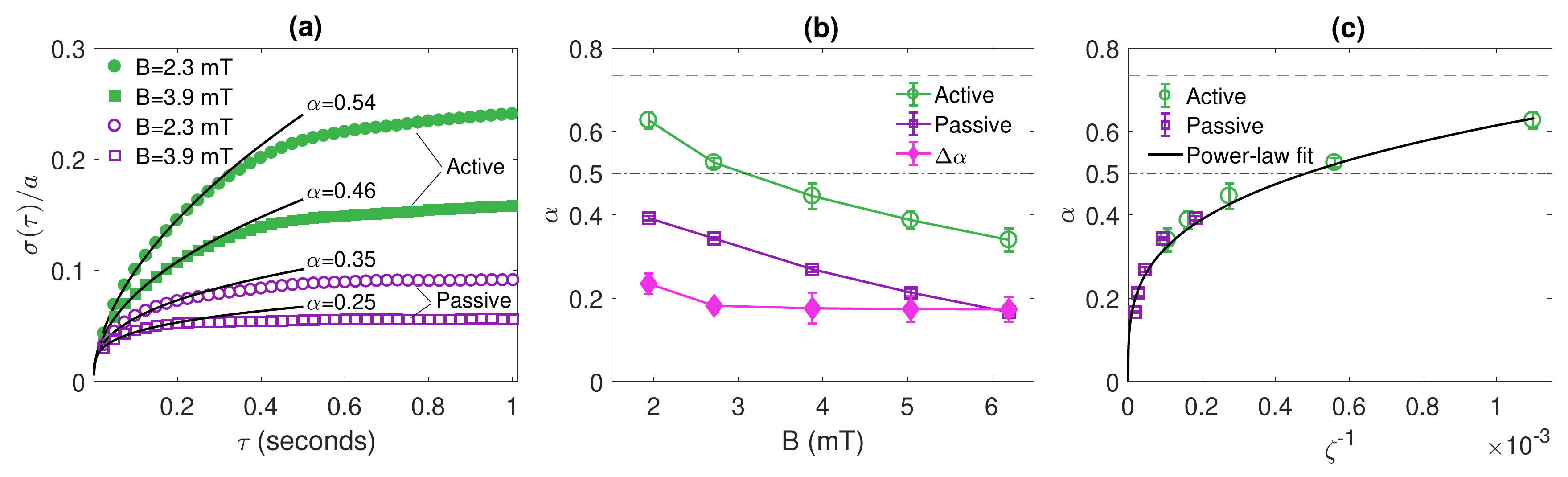}
\caption{Quantification of the internal chain dynamics in active and passive baths. (a) Root-mean-square (RMS) displacement $\sigma(\tau)$, as defined in Eq.~\eqref{eq:sigma} of a chain containing $N=7$ particles in active and passive baths for various magnetic field strengths $B$. Solid lines show fits to $\sigma(\tau) \propto \tau^\alpha$ in the short-time regime. (b) The fitted exponent $\alpha$ as a function of $B$. (c) $\alpha$ as a function of the inverse of the dimensionless coupling parameter $\zeta$, defined in Eq.~\eqref{eq:zeta}, using the effective temperature $T_{\rm eff} \approx 1760$ K for the active bath data. The horizontal dashed lines in (b) and (c) show the exponents for free particles ($N=1$) in passive (purple, dash-dotted line) and active (green, dashed line) baths. Error bars show one standard deviation  across independent experiments.}
\label{fig2}
\end{figure*}

The sample cell is a tube with a conical internal shape where a droplet of the sample solution is placed (Figs.~\ref{fig1}a-b). In order to minimize particle drift due to interfacial curvature, the amount of liquid is adjusted to form a flat liquid-air interface at the bottom of the sample volume, where the inner bottom diameter of the sample holder is $4\,{\rm mm}$. The paramagnetic particles in the solution (Microparticles PS-MAG-S2180, diameter $3.9\,{\rm \upmu m}$) sediment to this interface because of gravity, where they get absorbed. Using this sample geometry helps avoiding problems related to particles sticking to a solid surface; the much lower friction of the liquid-air interface furthermore enables assembling the magnetic chains using significantly lower magnetic field strengths.
 
The DC magnetic field is generated using a Helmholtz coil configuration (Fig.~\ref{fig1}a). Two solenoidal coils with an inner diameter of $3.4\,{\rm cm}$ are connected serially at a $3.5\,{\rm cm}$ separation. This configuration generates a highly homogeneous magnetic field across the sample, with less than $1\%$ variation over the entire diameter of the sample cell. Under the influence of the external magnetic field, the superparamagnetic particles become magnetized, and the resulting magnetic interactions between them lead to the formation of isolated chains of particles aligned with the field direction.
 
Motile \emph{Escherichia coli} bacteria were cultured following the protocol in Ref.~\cite{pincce2016disorder}. Briefly, wild-type \emph{E. coli} (strain RP437) were cultured overnight at $32^\circ{\rm C}$ in $50\,{\rm mL}$ Tryptone Broth ($1\%$ Tryptone, $0.5\%$ NaCl) on a rotary shaker ($180\,{\rm rpm}$). The saturated culture was diluted 50 fold into fresh medium and incubated again for $4\,{\rm h}$ at $32^\circ{\rm C}$ until the culture reached its mid-exponential growth phase (${\rm OD\,600\sim0.4}$) to ensure the motility of the bacteria. Next, the culture was washed and resuspended twice in motility buffer containing $10\,{\rm mM}$ potassium phosphate monobasic ($\text{KH}_2\text{PO}_4$), $0.1\,{\rm mM}$ EDTA (pH 7.0), $10\,{\rm mM}$ Dextrose ($\text{C}_6\text{H}_{12}\text{O}_6$), and $0.002\%$ Tween 20. The typical surface density of bacteria at the interface was $n \approx 3.5 \times 10^4\,{\rm cells\,mm^{-2}}$.

The sample was imaged using a custom-built microscope consisting of a $20\times$ objective lens (${\rm NA}=0.50$, ${\rm WD}=0.17\,{\rm mm}$) and a digital camera (Thorlabs DCC1645C). For the center-of-mass analysis, we recorded videos of isolated chains at 15 frames per second for at least 3 minutes. For the analysis of internal dynamics, videos were recorded at 40 frames per second in the active bath and 80 frames per seconds in the passive bath to properly capture the short-time regime.  A sample image of a chain in a bacterial bath is shown in Fig.~\ref{fig1}c; movies showing the chain dynamics in both active and passive baths can be found in Supplemental Material\dag. The videos were analysed with digital video microscopy \cite{crocker1996methods} and the chain dynamics were analysed in the internal coordinate system shown in Fig.~\ref{fig1}d.

\section{Results and discussion}

We begin by studying the internal chain dynamics as a function of the field strength $B = |\mathbf{B}|$, and how these are affected by the active noise induced by the bacterial bath. In Fig.~\ref{fig2}, for a chain containing $N=7$ particles, we analyse the root-mean-square (RMS) displacement $\sigma(\tau)$ perpendicular to the chain direction, defined by~\cite{silva1996fluctuation}
\begin{equation}\label{eq:sigma}
\sigma(\tau) = \langle \left[ h_i(t) - h_i(t+\tau) \right]^2 \rangle^{1/2}, 
\end{equation}
where $h_i(t)$ denotes the deflection of the $i^{\mathrm{th}}$ particle along the chain in the direction perpendicular to $\mathbf{B}$ at time $t$ (see Fig.~\ref{fig1}d), and angular brackets denote an average over all times $t$ and particles $i$. $\sigma(\tau)$ has previously been shown to exhibit a power-law behavior $\sigma(\tau) \propto \tau^\alpha$ for short times, before reaching a plateau value at longer times due to the effective confinement of each particle in the dipolar potential exerted by the rest of the chain particles \cite{furst2000dynamics, silva1996fluctuation}.
For a freely diffusing particle, $\alpha = 0.5$, while for a dipolar chain in a passive bath $\alpha < 0.5$ due to the hindered diffusion \cite{silva1996fluctuation}. In Fig.~\ref{fig2}a, we confirm this subdiffusive process for a chain in a passive bath at two different values of $B$ (purple curves): $\alpha$ decreases significantly as $B$ increases due to the larger strength of the dipolar interactions that hinder particle diffusion. In the active bath (green curves), the power-law behavior persists, but with a distinctly enhanced internal chain dynamics visible through a higher initial exponent for $\sigma(\tau)$. Furthermore, in spite of the hindered diffusion of each particle, the exponent reaches superdiffusive values ($\alpha > 0.5$) for low enough $B$-values ($B \approx 3 $ mT or less). 

Considering now the detailed dependence of $\alpha$ as a function of $B$ (Fig.~\ref{fig2}b), we observe  that the difference  $\alpha_{\mathrm{active}} - \alpha_{\mathrm{passive}}$ remains essentially constant as $B$ is varied; in other words, for small $\tau$, the ratio $\sigma_{\mathrm{active}}(\tau)/\sigma_{\mathrm{passive}}(\tau)$ is independent of $B$. This result indicates that the fluctuations due to the active bath qualitatively affect the chain dynamics in a similar way as those due to the passive bath, but with a higher noise magnitude. To further investigate the validity of this ``effective temperature'' picture of the internal chain dynamics, in Fig.~\ref{fig2}c we plot $\alpha$ as a function of the inverse of the  dimensionless magnetic coupling strength 
\begin{equation}\label{eq:zeta}
\zeta \equiv B^2 a^3 / (\mu_0 k_{\rm B} T),
\end{equation}
where $a = 1.95\,{\rm \upmu m}$ is the particle radius , $T$ is the absolute temperature and $\mu_0$ is the vacuum permeability. In the passive bath, $\zeta$ is proportional to the ratio between the dipolar coupling and the thermal energy, modulo an unknown factor of magnetic susceptibility. In the active bath, we instead use $T = T_{\rm eff}$ as a free parameter characterising the active bath. Using $T_{\rm eff} \approx 6T_0$, where $T_0=294\,{\rm K}$ is the laboratory temperature, we obtain an excellent collapse of the two experimental curves using a common power law $\alpha \propto \zeta^{-\beta}$, with $\beta \approx 0.25$, indicating that the effective temperature picture remains valid for relatively stiff chains (\emph{i.e.}, high $\zeta$). For flexible chains, however, $\alpha$ has a strict upper bound for the free diffusion value $\alpha = 0.5$ in the passive bath (dashed purple line in Figs.~\ref{fig2}b-c), while it continues to increase beyond $0.5$ in the active bath. The active curve naturally has an upper theoretical bound at $\alpha = 1$, corresponding to ballistic motion; however, this regime is not experimentally accessible due to fast breakup of the chains at these low $B$ values.

\begin{figure*}
\includegraphics[width=2.0\columnwidth]{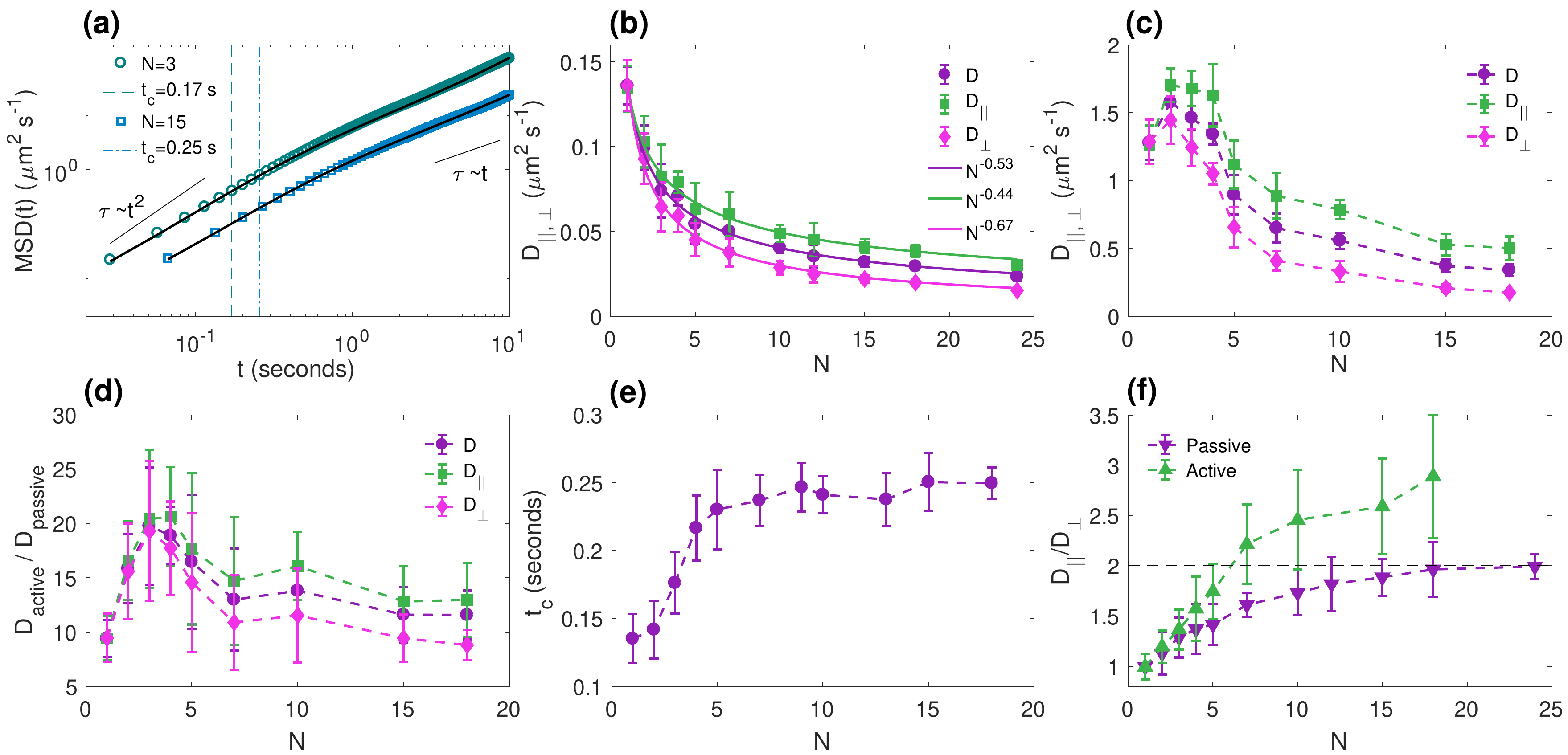}
\caption{Quantification of the anisotropic center-of-mass dynamics. (a) The total mean-square displacement (MSD) of chains of two different lengths ($N=3$ and $N=15$) in an active bath. Dashed and dash-dotted vertical lines indicate the crossover time $t_{\rm c}$ between the superdiffusive and diffusive regimes for $N=3$ and $N=15$, respectively, and solid lines are fits to Eq.~\eqref{MSD_active}. (b, c) Diffusion coefficients as a function of chain length $N$, in (b) a passive bath (motility buffer) and (c) in an active bacterial bath. (d) Ratio between the diffusion coefficients in active and passive baths as a function of $N$. (e) Measured crossover time $t_{\rm c}$ as a function of $N$. (f) Ratio between the diffusion coefficients in directions parallel ($D_{\parallel}$) and perpendicular ($D_{\perp}$) to $\mathbf{B}$, showing that the ratio in bacterial bath is somewhat larger than  the equilibrium (passive) system. Dashed line shows the theoretical limit for the ratios in passive bath. Magnetic field amplitude: $B=3.1$ mT. Error bars indicate one standard deviation  across independent experiments.}
\label{fig3}
\end{figure*}

In Fig.~\ref{fig3}, we characterize the anisotropic center-of-mass dynamics of the chain. In Fig.~\ref{fig3}a, we show typical mean square displacement (MSD) curves for chains of two different lengths in the active bath. We can clearly distinguish a short-time superdiffusive regime and a long-time diffusive regime, with the crossover occurring at a characteristic time $t_{\rm c}$ (dashed vertical lines in Fig.~\ref{fig3}a), as for spherical active colloids \cite{howse2007self}. In the active bath, we measure $t_{\rm c}$ and the diffusion coefficient $D$ for both the parallel and perpendicular components by fitting experimental data to the theoretical relation for the MSD of an active particle in 2 dimensions~\citep{howse2007self, bechinger2016active, eisenstecken2016conformational}: 
\begin{equation} \label{MSD_active}
\langle
|\mathbf{r}(t)-\mathbf{r}(0)|^2
\rangle 
=( 4D_0 +2t_{\rm c} v_0^2)t +2t_{\rm c}^2 v_0^2(e^{-t/t_{\rm c}}-1),
\end{equation}
where $\mathbf{r}(t)$ is the center-of-mass position at time $t$,  $D_0$ is the Stokes-Einstein diffusion coefficient, $t_{\rm c}$ is the crossover time between ballistic and diffusive motion, and $v_0$ is a typical center-of-mass velocity induced by the active noise. 

Fig.~\ref{fig3}b shows that the total center-of-mass diffusion coefficient of our short semiflexible chains in the passive bath fits well with a Zimm-like power-law scaling $D \propto N^{-\gamma}$~\cite{zimm1956dynamics}. While Zimm \cite{zimm1956dynamics} solved analytically the equations in his model at the case of large particle number N, our experimental observation is in agreement with previously reported results for short semiflexible chains in a passive bath~\cite{biswas2017linking}. Our measured exponent $\gamma \approx 0.53$ is somewhat smaller than the theoretical value $\gamma \approx 0.6$ predicted for a 3-dimensional chain in the limit $N \rightarrow \infty$~\citep{doi1988theory}, probably due to the quasi-2-dimensional dynamics and the fairly short chains considered. We also find that both $D_{\parallel}$ and $D_{\perp}$ follow similar power-law scaling, but with somewhat different exponents.

The corresponding diffusion coefficients in the active bath, obtained by fitting the MSDs to Eq.~\eqref{MSD_active}, are shown in Fig.~\ref{fig3}c. Clearly, both $D_{\parallel}$ and $D_{\perp}$ are about an order of magnitude larger than the corresponding thermal diffusion coefficient (see Fig.~\ref{fig3}d), yielding a value between $T_{\rm eff} \approx 2650\,{\rm K}$ ($N = 1$) and $T_{\rm eff} \approx 5880\,{\rm K}$ ($N = 3$), assuming that $T_{\rm eff} / T_0 = D_{\mathrm{active}} / D_{\mathrm{passive}}$. For $N=7$, the case also studied in Fig.~\ref{fig2}, we find $T_{\rm eff} \approx 3650\,{\rm K}$, \emph{i.e.}, about twice as large as the value obtained from the internal chain dynamics ($T_{\rm eff} \approx 1760\,{\rm K}$). This once again highlights the fact that transient short-time properties (such as the internal chain dynamics analyzed in Fig.~\ref{fig2}) often yield different dynamics than the steady-state long-time properties, such as the center-of-mass diffusion, even when they can individually be analyzed in terms of an effective temperature, thus showing the limitations of simplified thermodynamic descriptions of active systems.

We furthermore observe that both the parallel and perpendicular active diffusion coefficients are \emph{non-monotonic} in the chain length $N$,
with a maximum diffusion coefficient for $N=2$. A similar effect has previously been observed for spherical tracers in an \emph{E. coli} suspension \cite{patteson2016particle}, where the maximum in $D$ was observed for particle diameters of $\sim 2-10\,{\rm \upmu m}$, comparable to our maximum diffusion at a chain length of $\sim 8\,{\rm \upmu m}$. Despite the different geometry of spheres and chains, this agreement indicates that the size-dependent behaviour is a direct consequence of the active bath properties, such as the bacteria size and concentration, rather than of the precise particle geometry.  (Note, furthermore, that this maximum in diffusivity is different from that observed in Ref.~\cite{shin2017elasticity}, which occurs for much longer chains, and is explained by shape fluctuations due to chain flexibility.) 

In Fig.~\ref{fig3}e, we show that this non-monotonicity is caused by an initial increase of the crossover time $t_{\rm c}$ as a function of chain length $N$, as, for a constant $v_0$, $D$ will increase monotonically with $t_{\rm c}$ according to Eq.~\eqref{MSD_active}. Since $v_0$ monotonically decreases with particle size due to the increased friction of the chain, $D$ develops a peak at an intermediate value of the particle size.

This increase in $t_{\rm c}$ with chain length is also in accordance with what was measured previously for spherical particles in Ref. \cite{patteson2016particle}. This size-dependent crossover time can have several explanations, for example effects due to nonlinearities in the swimmer flow fields leading to size-dependent tracer advection in accordance with Fax\'en's law, or changes in the bacterial trajectories due to the presence of the chain leading to more persistent swimmer-chain scattering events for longer chains. 

In spite of the greatly enhanced values compared to the passive case of both the parallel ($D_{\parallel}$) and perpendicular ($D_{\perp}$) components of the diffusion coefficient, we find that their \emph{ratio} $D_{\parallel} / D_{\perp}$ (Fig.~\ref{fig3}f) remains above 1, similar to that in the thermal (passive) system, although with a somewhat higher ratio than in the passive case for large $N$: In the passive case, our results are in accordance with the bulk prediction $D_{\parallel} / D_{\perp} \rightarrow 2$ for large $N$~\cite{dhont1996introduction}, while in the active bath the ratio approaches a value of $D_{\parallel} / D_{\perp} \sim 3$ for the range of chain lengths considered here. This finding is qualitatively different from the anisotropic diffusion measured for free ellipsoidal tracer particles in an \emph{E. coli} suspension in Ref.~\cite{peng2016diffusion}, where a \emph{reversal} in the anisotropy was measured, \emph{i.e.}, $D_{\parallel} / D_{\perp} < 1$; we will discuss the possible origins of these differences in more detail below. 

\begin{figure}
\includegraphics[width=1\columnwidth]{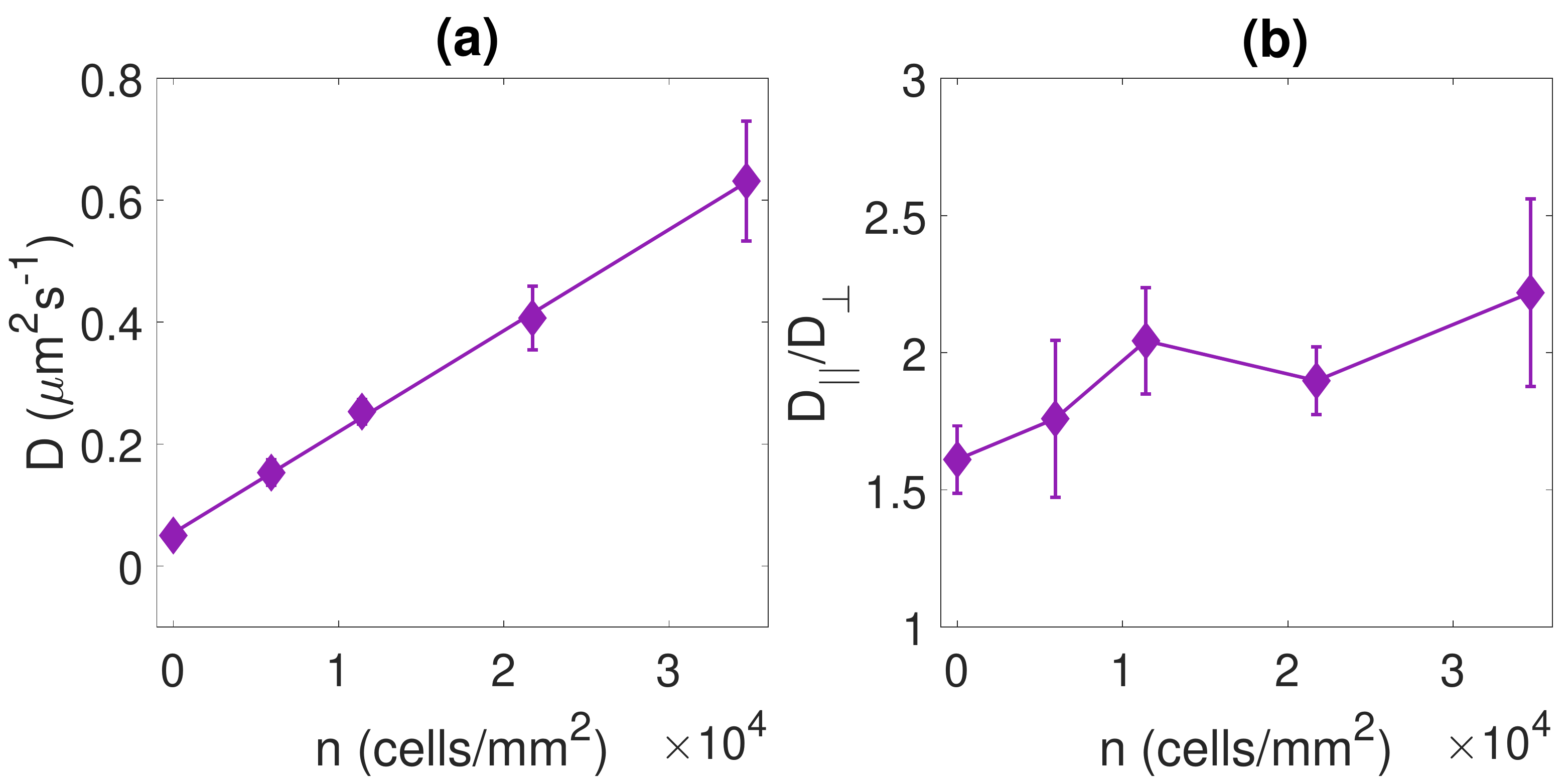}
\caption{Density dependence of the center-of-mass dynamics. (a) The total diffusion coefficient $D$ and (b) the ratio  $D_{\parallel}/D_{\perp}$,  as a function of the two-dimensional bacterial density $n$ at the liquid-air interface for $N=7$. Magnetic field amplitude: $B=3.1$ mT. Error bars show one standard deviation across independent experiments.}
\label{fig4}
\end{figure}

In Fig.~\ref{fig4}a, we investigate the dependence of the center-of-mass dynamics on the bacterial concentration $n$. These results reproduce the linear dependence of $D$ on $n$ previously observed for spherical tracers \cite{jepson2013enhanced,mino2011enhanced,leptos2009dynamics}, indicating that the enhanced diffusion is a single-particle effect and not due to collective or correlated motion between bacteria, in which case a superlinear behavior is to be expected~\cite{stenhammar2017role}. Furthermore, in Fig.~\ref{fig4}b, we observe that the measured value of the ratio $D_{\parallel} / D_{\perp}$ is a weakly increasing function of the bacterial concentration, again contrasting the results of Ref.~\cite{peng2016diffusion}, where $D_{\parallel} / D_{\perp}$ was found to be a strongly decreasing function of $n$ for ellipsoidal particles in the same range of bacterial concentrations~\footnote[3]{The two-dimensional bacterial concentration in Ref.~\cite{peng2016diffusion}, assuming one monolayer on each interface, ranges between $6.0  \times 10^3\,{\rm mm^{-2}} \leq n \leq 2.4 \times 10^5\,{\rm mm^{-2}}$. The two-dimensional bacteria concentration in our experiments is approximately $n \approx 3.5\times 10^4\,{\rm cells\,mm^{-2}}$.}.
  
As mentioned above, our self-assembled chains are very stiff due to the external magnetic field, and exhibit small fluctuations about their rigid rod limit. In other words, in our system the radius of gyration \citep{doi1988theory} is very close to $Na$ where $a$ is the radius of the particles.  Therefore, the internal chain fluctuations are unlikely to be responsible for this  different behavior.
While the difference in particle geometry (particle chain versus ellipsoid) between the studies might explain moderate quantitative differences, they seem unlikely to lead to a qualitative change in behavior. Instead, this difference is probably due to differences in experimental conditions and setups. First of all, the linear dependence of $D$ on $n$ (Fig.~\ref{fig4}) indicates that our system does not feature collective behavior, while the anomalous relationship between $D_{\parallel}$ and $D_{\perp}$ in Ref.~\cite{peng2016diffusion} occurred in the regime where collective motion (``active turbulence'') was observed. While we work in similar concentration regimes as in Ref.~\cite{peng2016diffusion}, the presence or absence of such collective motion can also be strongly dependent on the details of the experimental geometry used. Secondly, the results in Ref.~\cite{peng2016diffusion} were obtained in a free-standing soap film (approximately 15 $\mu$m thick). Since the ellipsoidal tracers in Ref.~\cite{peng2016diffusion} were large enough to effectively be confined between the film interfaces (the long axis had a length of $\sim 28~\mu$m), direct collisions between bacteria and tracers were facilitated compared to the case of a three-dimensional suspension~\cite{berke2008hydrodynamic, shum2015hydrodynamic}. In contrast, in the experimental geometry used in our study, the bacteria are free to swim in three dimensions (although they tend to accumulate near the interface), while the chain is effectively confined to the liquid-air interface by gravity. Thus, the number of direct collisions will be lower, and the advection of the chain will instead be dominated by long-range hydrodynamic scattering events; such hydrodynamic scattering has previously been shown to dominate the observed enhanced diffusion of spherical tracer particles in three-dimensional \emph{E. coli} suspensions~\cite{jepson2013enhanced}. Changing this subtle balance between long-ranged hydrodynamic advection and direct collisions may strongly alter the chain dynamics in a non-trivial way.

\section{Conclusions}

In this study, we have investigated the internal and center-of-mass dynamics of a self-assembled colloidal chain composed of magnetic particles in an active bath containing swimming \emph{E. coli}. We found that both the internal dynamics and the center-of-mass diffusion are strongly enhanced, by an order of magnitude or more, compared to those in the passive bath. Our results are furthermore broadly consistent with an ``effective temperature'' description, although with different magnitudes for the short-time internal chain dynamics and the long-time center-of-mass dynamics, thus highlighting the limitations of such simplified thermodynamic descriptions of active systems. We furthermore reproduced the anomalous, non-monotonic size dependence of the diffusion coefficient previously observed for spherical particles~\cite{patteson2016particle}, and showed that it is due to an initial increase of the crossover time $t_{\rm c}$ between the superdiffusive and diffusive regimes in the mean-square displacement. In qualitative contrast to previous results for anisotropic particles in \emph{E. coli} suspensions \cite{peng2016diffusion}, we however found that the ratio $D_{\parallel} / D_{\perp}$ between the components of $D$ parallel and perpendicular to the direction of the applied magnetic field is similar to, although slightly larger than, the equilibrium (passive) case within the full range of bacterial densities considered. While the origin of this qualitative difference is not obvious, it might be due to the absence or presence of collective motion among bacteria, as well as subtle but important differences in the experimental geometries altering the balance between direct collisions and long-range hydrodynamic advection. Our results thus indicate that the observed tracer dynamics will depend sensitively on the experimental characteristics of the ``active bath'' in a given system, and that this balance is strongly dependent on the details of the experimental geometry. In particular, one cannot expect 2-dimensional and 3-dimensional active matter models to be qualitatively equivalent, since the role of direct particle collisions will be more important in two dimensions than in three dimensions, at the expense of hydrodynamically induced forces. Our observations thus call for further experiments in well-controlled experimental geometries to elucidate the complex dynamics of isotropic and anisotropic objects in active baths.

\section{Acknowledgments}
Helpful discussions with Alexander Morozov and Sabareesh K.P. Velu are kindly acknowledged.  
This work was supported by the Scientific and Technological Research Council of Turkey (TUBITAK 215E198). 
JS is funded by the Knut and Alice Wallenberg foundation (KAW 2014.0052) and the Swedish Research Council
(2019-03718).

\balance

\bibliography{Bibliography}
\bibliographystyle{rsc}

\end{document}